\documentclass[final,authoryear,3p,times]{elsarticle}

\usepackage{ecrc,natbib} 
\usepackage{mathtools}
\usepackage{listings}
\usepackage{multirow}
\usepackage{xcolor}

\volume{00}

\firstpage{1}

\journalname{Astronomy and Computing}

\runauth{Desai et al.}

\jid{procs}

\jnltitlelogo{Astronomy and computing}

\usepackage{amssymb}
\usepackage[figuresright]{rotating}

\definecolor{backcolour}{rgb}{0.95,0.95,0.92}

\begin{document}

\begin{frontmatter}

\title{Detection and Removal of Artifacts in Astronomical Images}

\author[lmu,exc]{Shantanu Desai}
\ead{shntn05@gmail.com}
\author[lmu,exc,mpe]{Joseph J. Mohr\,}
\author[iap]{Emmanuel Bertin}
\author[lmu]{Martin K\"ummel}
\author[lmu]{\\ Michael Wetzstein}

\address[lmu]{Faculty of Physics, Ludwig-Maximilians-Universit\"{a}t, Scheinerstr.\ 1, 81679 Munich, Germany}
\address[exc]{Excellence Cluster Universe, Boltzmannstr.\ 2, 85748 Garching, Germany}
\address[mpe]{Max Planck Institute for Extraterrestrial Physics, Giessenbachstr.\ 85748 Garching, Germany}
\address[iap]{Institut d'Astrophysique de Paris, 98 bis boulevard Arago, 75014 Paris, France}
\begin{abstract}

Astronomical images from optical photometric surveys are typically contaminated with transient artifacts such as  cosmic rays, satellite trails and scattered light.  We have developed and tested an algorithm that removes these artifacts using a deep, artifact free, static sky coadd image built up through the median combination of point spread function (PSF) homogenized, overlapping single epoch images.  Transient artifacts are detected and masked in each single epoch image through comparison with an artifact free, PSF-matched simulated image that is constructed using the PSF-corrected, model fitting catalog from the artifact free coadd image together with the position variable PSF model of the single epoch image.  This approach works well not only for cleaning single epoch images with worse seeing than the PSF homogenized coadd, but also the traditionally much more challenging problem of cleaning single epoch images with better seeing.  In addition to masking transient artifacts, we have developed an interpolation approach that uses the local PSF and performs well in removing artifacts whose widths are smaller than the PSF full width at half maximum, including cosmic rays, the peaks of saturated stars and bleed trails.  We have tested this algorithm on Dark Energy Survey Science Verification data and present performance metrics. More generally,  our algorithm can be applied to any survey which images the same part of the sky multiple times.
\end{abstract}


\end{frontmatter}

\section{Introduction}

In the last two decades, many  optical wide-field photometric surveys using new, state of the art  multi-megapixel cameras have started taking data and  mapping out large portions of sky. A pioneering effort on this front has been from the Sloan Digital Sky Survey (SDSS)~\citep{SDSS}.  In the last few years many surveys deeper than SDSS have commenced operations. These  include DES~\citep{DES05,DES16}, KiDS~\citep{KIDS13}, and  HSC~\citep{HSC12}.   Within a decade, LSST will start taking data mapping half the sky with unprecedented depth~\citep{LSST}. The main science goal of these surveys is to constrain cosmological parameters and provide insight into the underlying causes of the cosmic acceleration and the characteristics of dark matter and inflation.  These studies rely on techniques that require observations of Type 1a supernovae, galaxy clusters, weak lensing, and baryon acoustic oscillations, as well as measurements within our own galaxy.  These homogeneous, wide field survey datasets enable a plethora of astrophysical studies and studies of structure formation in general.
 
To achieve these myriad science goals, it is important to identify and remove both transient and persistent artifacts from the images prior to production of the science-ready catalogs. The transient artifacts are those features such as cosmic rays, satellite trails, scattered light and artifacts due to malfunctioning CCDs that appear in only one of a series of observations of the same sky location.  The artifacts that persist among an ensemble of observations of the same sky location include saturated bright stars, bleed-trails, and diffraction spikes.  These artifacts can be detected and characterized as objects, contaminating the science catalogs produced in these surveys.  Moreover, these artifacts can in principle affect photometric and astrometric calibration, degrading the catalogued information from surrounding objects.

In creating a deep, coadded image of a sky location one can simply apply an outlier rejection routine or even median filter the input single epoch images \citep[e.g.][]{Gruen14}.  While this works well in identifying transient artifacts that appear in empty, sky dominated regions of the sky, it fails in identifying artifacts that lie on or near real objects because of the image PSF variations over time that are generic in the acquisition of large survey datasets.  These seeing variations generically result in large apparent variations in star and galaxy morphologies that then can be incorrectly identified as artifacts.  Thus, a method that works robustly within a dataset that includes multiple visits to each location on the sky under different imaging conditions is needed. Moreover, many surveys are now focused on cataloging the single epoch images-- using the appropriate PSF for each single epoch image and producing a single combined catalog-- rather than first producing PSF homogenized coadd images where the noise is necessarily correlated among neighboring pixels and then cataloging them.  Such an approach requires that the artifact be detected and masked within the full single epoch imaging dataset.  

It is often convenient also to remove the artifacts from the single epoch images once they have been identified.  One motivation for this is that in those surveys that catalog using the coadd images, it has now become common to transform to a common PSF before coaddition.  This approach avoids spatial discontinuities in the effective PSF within the coadd image that then make it impossible to extract accurate PSF corrected object photometry or even to simply identify unresolved objects \citep[see, e.g.,][]{Desai12}.  An efficient way of doing this involves application of a spatially varying convolution kernel and fast Fourier transform techniques.  In such an approach a single pixel artifact becomes a lower amplitude artifact with the scale of the homogenizing kernel, which is characteristically the size of the PSF.  In the case of artifacts whose dimension is smaller than the width of the PSF, interpolation to remove them works quite well with only a slight increase in noise.  For artifacts that are comparable to or larger than a PSF there is no way to recover the information loss in the image and there is no real gain in removal of the artifact.

While significant effort has already been invested in the development of algorithms to detect  cosmic rays in single epoch images~\citep{Rhoads00,vandokkum01,Bertin01,Farage05,Ipatov07}, these new survey datasets and the availability  of large scale computing facilities enable new, more robust techniques to be applied.  Most of these algorithms suffer from some weaknesses, and either do not correctly identify all the cosmic rays or sometimes identify faint objects as artifacts.  Also these algorithms tend to have difficulty in detecting cosmic rays that lie within real astrophysical objects.  To remove satellite tracks from individual images, Hough transform techniques are often used~\citep[e.g.,][]{Hough}. To the best of our knowledge, there is currently no automated algorithm to detect scattered light within single epoch astronomical images, and sometimes these are manually identified through painstakingly scanning each image~\citep{Jarvis}.  

In this paper, we describe a new method to detect and remove artifacts in astronomical images.  We briefly describe the data management system used for this test in Section~\ref{sec:cosmodm}.  Section~\ref{sec:algorithm} contains a description of our artifact detection and masking algorithm, which operates autonomously on transient artifacts.  It relies on (1) a dataset that contains multiple visits to the same portion of the sky, (2) accurate modeling of the position variable PSF on single epoch images, (3) the construction of deep, PSF homogenized artifact-free images, (4) model fitting cataloging of that image and (5) the production of position variable PSF convolved simulated images using these model fitting catalogs and PSF models.  Our approach is notionally related to the widely used {\tt Drizzle} algorithm \citep{Drizzle02}, which is applied in the processing of HST data but operates within a context of uniform, space-based imaging and does not rely on PSF corrected model fitting.  Section~\ref{sec:removal} contains a description of our techniques for removing artifacts, which, depending on the width of the artifact, employ either interpolation or replacement using a PSF convolved object model constrained by surrounding, uncontaminated pixels.   Section~\ref{sec:application} describes a test of our algorithms using science verification data from the Dark Energy Survey and a demonstration of the depth improvements that accompany the improvements in image quality.  In Section~\ref{sec:discussion} we discuss the shortcomings of our algorithm and future improvements, and we present our conclusions in Section~\ref{sec:conclusions}.

\section{Input Data Preparation with CosmoDM}
\label{sec:cosmodm}

The artifact removal tools we describe here have been developed as part of the Cosmology Data Management system (CosmoDM).  CosmoDM has been developed at Ludwig-Maximilians-Universit\"at (LMU) in Munich since 2011;  it arose from a development version of the  Dark Energy Survey Data Management system~\citep[DESDM;][]{Ngeow06,Mohr08,Mohr12}.  CosmoDM and its precursor have been applied to a variety of data including those from the Mosaic-2 and DECam imagers on the Blanco telescope as well as data from Pan-Starrs, CFHT MegaCam, SOAR, and WFI to support the optical confirmation and photometric redshift measurement of galaxy cluster candidates detected through both Sunyaev-Zeldovich effect surveys \citep{Staniszewski09,Song12,Desai12,Liu15,Desai15} and X-ray surveys \citep{Suhada12}.  The science codes and quality assurance framework from CosmoDM serve as a prototype for the ground-based external dataset calibration pipelines needed for the Euclid satellite mission~\citep{Euclid11}.  

CosmoDM is an automated system that includes, among other components, pipelines for processing and calibrating single epoch images and for building and cataloging coadd images.  In our system we differentiate between single epoch exposures (raw data from the telescope), single epoch images (the detrended and calibrated CCD-sized images-- $\sim$10$'\times$20$'$ for DECam) and deeper, stacked coadd images that are built from all the available single epoch images in a particular sky location. Typically these CosmoDM pipelines are run in a fashion where one wishes to build deep, coadd images in a particular sky region or tile (typically $1^\circ \times 1^\circ$  sky regions); the system first finds all single epoch exposures that overlap the coadd tile, prepares the associated single epoch images and then combines them into a coadd image for each photometric band. The single epoch processing and calibration starts from the raw exposures and ends with astrometrically calibrated, fully flattened and detrended single epoch images, position variable PSF models and associated PSF corrected model fitting catalogs. The detrending corrections include crosstalk corrections, overscan subtraction, bias corrections, flat fielding, photometric flattening and astrometric calibration.  Noise is tracked for each pixel in an associated weight map and bad pixel maps (BPMs) encode any special actions taken on particular pixels.  Astrometric calibration is done with {\tt SCAMP}~\citep{Scamp06}, where for DECam~\citep{Decam15} we employ a third-order polynomial distortion correction and use a DECam extracted distortion catalog to constrain the field distortions and the 2MASS~\citep{Twomass06} catalog for absolute astrometric calibration, respectively.  For the PSF corrected model fitting catalogs, we use PSF models extracted using {\tt  PSFEx}~\citep{Bertin11} together with {\tt SExtractor}~\citep{Bertin96}.  We use the model fitting capabilities added to {\tt SExtractor} as part of the DESDM development program and extract more than 100 morphological and photometric parameters for every detected object.

Once the basic single epoch image processing and calibration is complete, we use the catalogs to determine relative photometric zeropoints among the ensemble of images within each band and refine the astrometric solutions within the single epoch images by a second run of {\tt SCAMP} that uses the entire ensemble of overlapping images to constrain the astrometric distortion parameters.  Typical astrometric accuracy after this stage is $\sim$20~milli-arcsec.  These astrometrically and photometrically calibrated images are then used to build PSF homogenized coadds. To create PSF-homogenized coadds, we first convolve the single epoch images with a PSF transformation kernel from {\tt PSFex} that allows us to bring all images to a single, constant Moffat PSF model. We typically homogenize to the median PSF of all the single epoch images contributing to a single photometric band.  These PSF homogenized single epoch images are then background subtracted and combined using {\tt SWarp}~\citep{Bertin02}.  The associated weight maps corrected for the pixel correlations introduced by homogenization are also combined into an associated coadd image weight map.  With median combine this process produces a transient artifact free model of the light distribution in the sky convolved with a precisely known Moffat PSF-- even prior to single epoch image masking.  We catalog the PSF homogenized coadds using model-fitting photometry by running {\tt SExtractor} in dual image mode, with a $\chi^2$ detection image~\citep{Szalay99} constructed from a combination of the coadd images (typically $i$ and $z$-band).  The absolute photometric calibration is done using stellar locus regression~\citep{Desai12} on the multiband catalogs obtained from the coadd images.  

The single epoch images together with the homogenized coadd image and associated catalog described above form the key inputs to our masking algorithm.

\begin{figure}
\begin{center}
\includegraphics[width=0.65\columnwidth]{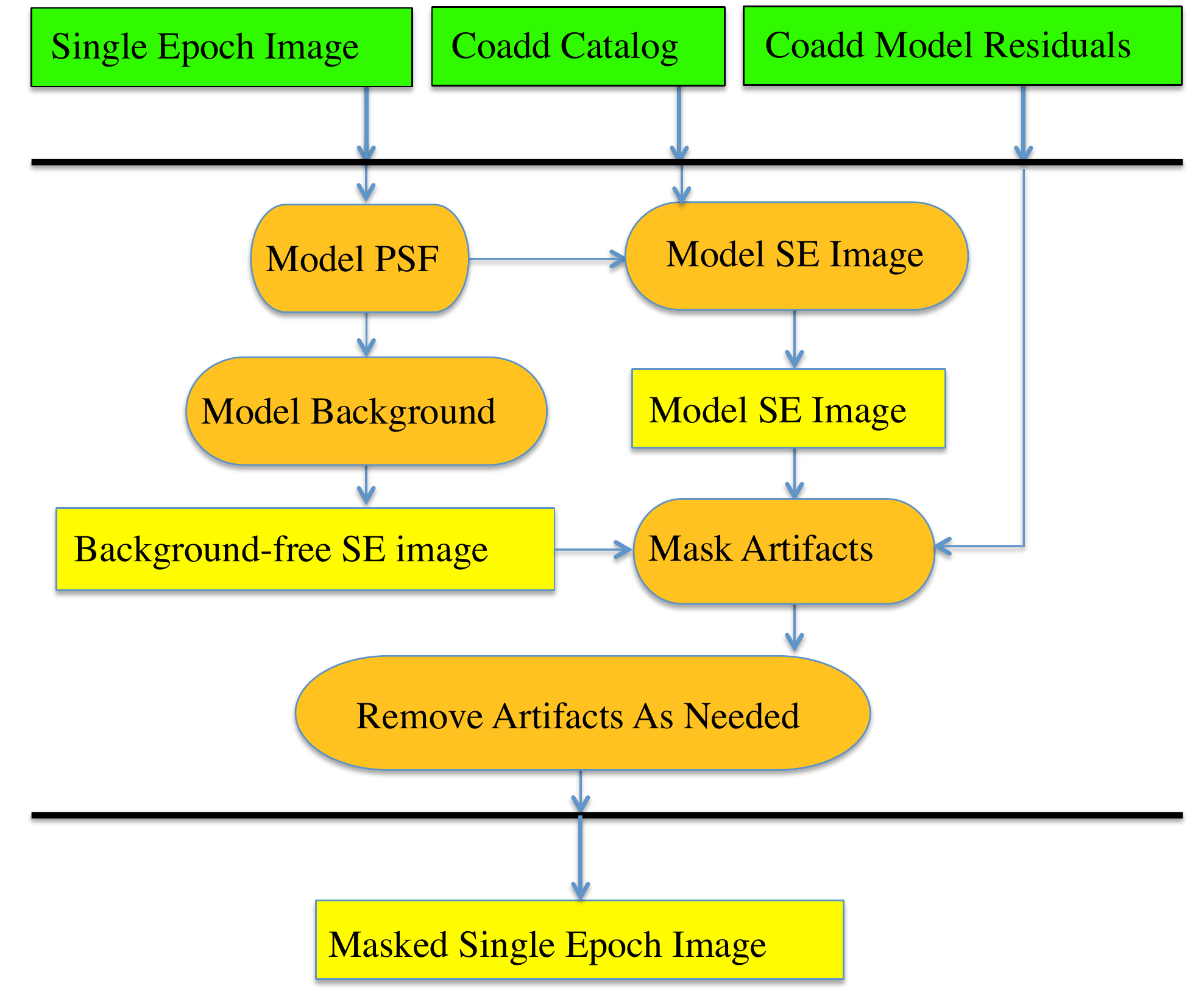}
\end{center}
\vskip-0.2in
\caption{Flow chart of the masking algorithm as applied to a single epoch image. Inputs are marked in green, actions are orange, and data products are in yellow.   Section~\ref{sec:algorithm} contains a description of each masking step, and the artifact removal is described in Section~\ref{sec:removal}.}
\label{fig:flowchart}
\end{figure}

\section{Algorithm to Detect Transient Artifacts}
\label{sec:algorithm}
We provide a conceptual overview of the method in Section~\ref{sec:method}, review the input data products in Section~\ref{sec:inputs}, describe the model light distribution produced for each single epoch image in Section~\ref{sec:model} and finally present the masking criterion in Section~\ref{sec:masking}.  A flow chart of our processing appears in Figure~\ref{fig:flowchart}.

\subsection{Overview of Method}
\label{sec:method}

The masking procedure requires a pixel by pixel comparison of each single epoch image with a high fidelity, artifact free and PSF matched model image.  The single epoch model image is produced from the single epoch, spatially varying PSF model together with the PSF corrected model fitting catalog extracted from the artifact free, PSF homogenized coadd image described in Section~\ref{sec:cosmodm}.  As described earlier, the coadd image is the deep image produced through the combination of multiple single epoch images overlapping a particular part of the sky.  All pixel differences between the single epoch image and associated single epoch model image that lie above a signal to noise threshold are then flagged.  To reduce the false positives introduced by model fitting failures in the coadd image, we veto flagged pixels in the single epoch image in areas where the difference between the coadd image and the coadd model image exceed a signal to noise threshold.  The coadd model image reflects the sum of the light distribution of all the objects in the coadd catalog convolved with the homogenized coadd PSF.

This method performs well everywhere that the coadd catalog provides an accurate description of the observed light distribution in the coadd image.  As we discuss more below, failures occur where the {\tt SExtractor} PSF corrected model fitting fails, and in general this is only within a small fraction of the image pixels ($<$0.5\% of pixels; see discussion in Section~\ref{sec:discussion}).  Ongoing improvements to {\tt SExtractor} model fitting and other similar codes will further reduce this fraction.  In general, transient artifacts such as cosmic rays, satellite trails and scattered light are detected successfully whether or not they overlap true sky objects.

Within the subsections below we describe the different elements of the masking algorithm data flow as outlined in Figure~\ref{fig:flowchart}.

\subsection{Required Input Data Products}
\label{sec:inputs}

There are three data products required to support the masking of the single epoch images.  All are identified in green boxes at the top of Figure~\ref{fig:flowchart}, and we provide a detailed description of each below.

\begin{enumerate}

\item The processed and calibrated {\it single epoch image} produced as described in Section~\ref{sec:cosmodm}.  Within CosmoDM each single epoch image consists of three header data units: (1) the calibrated image, (2) the bad pixel mask (BPM), and (3) the inverse variance weight map. The BPM contains an image map where each pixel contains the bitwise encoded reason that a pixel has been flagged (if at all).  Bits are assigned to track saturation, bad columns, cosmic rays, satellite trails and scattered light.  Saturated pixels and bad columns are marked already prior to the masking pipeline, where the bad pixel map is created using outliers in dome flats and bias corrects to identify dead or hot columns.  In general, the weight map values for pixels that have been flagged are set to zero to ensure they are not used in coaddition or cataloging.  

\item  The {\it coadd catalog} of PSF corrected model fitting components extracted from PSF homogenized, median combine coadd images to which the single epoch image contributed.  In general, a single epoch image overlaps more than one neighboring coadd tile (i.e. sky region), and so typically more than one coadd catalog is involved.  Within CosmoDM the homogenized coadd and the associated PSF corrected catalog are created as described in Section~\ref{sec:cosmodm}, but the required catalog could also be produced, in principle, through simultaneous PSF-corrected model fitting to the ensemble of overlapping single epoch images.

\item The {\it coadd model residuals} image for each homogenized coadd image, which identifies the pixel by pixel deviations in units of signal to noise of the PSF corrected model image from the original coadd image.  This data product provides pixel level information on exactly where the PSF corrected model fitting components listed above fail to accurately reproduce the observed light distribution.  This image is obtained in CosmoDM by running {\tt SExtractor} on the homogenized coadd image with the option {\tt -CHECKIMAGE\_TYPE -MODELS} and then scaling this image by the square root of the coadd image weight map, which contains the inverse noise variance of each pixel.  

\end{enumerate}

\begin{figure}
\begin{center}
\includegraphics[width=0.33\columnwidth]{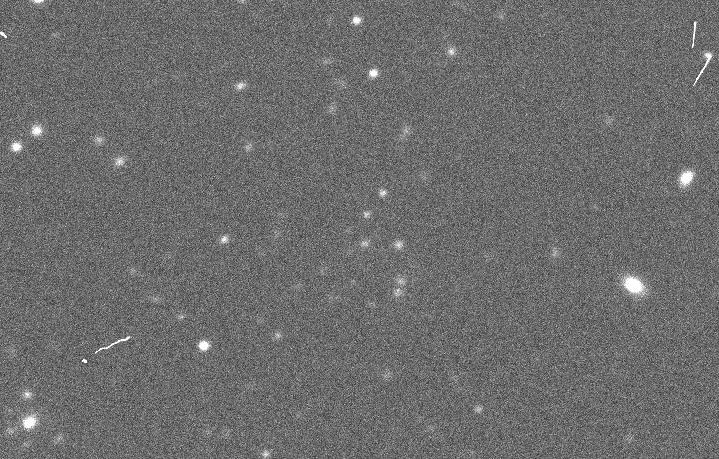}
\includegraphics[width=0.33\columnwidth]{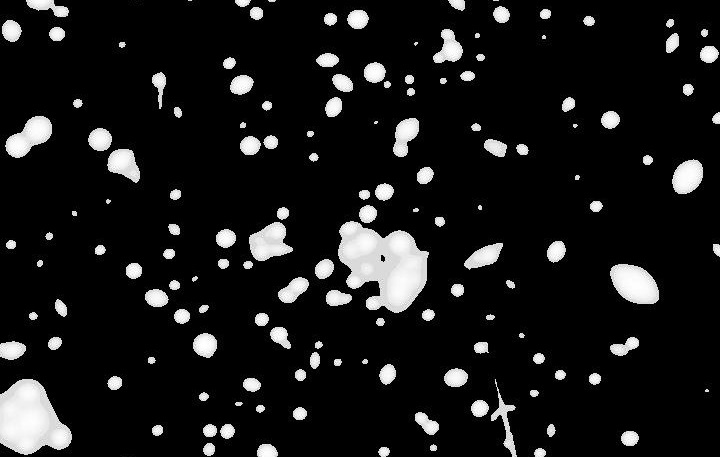}
\includegraphics[width=0.33\columnwidth]{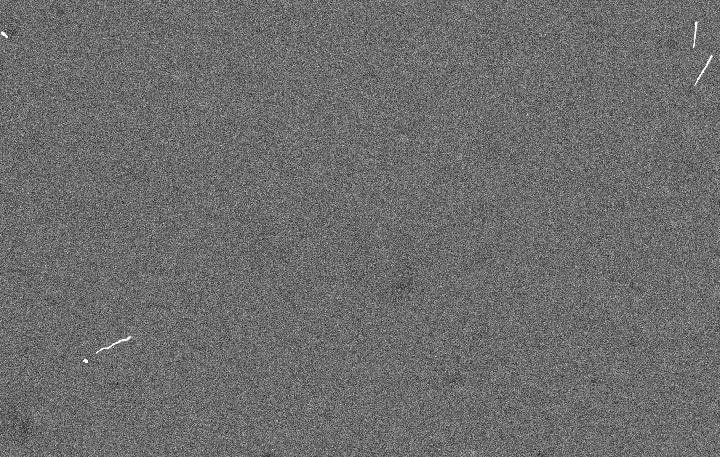}
\end{center}
\vskip-0.2in
\caption{DES-SV image (\#179097, CCD 27) cutout (left; $720 \times 460$ pixels) with corresponding model image (middle) and deviation image (right), which is used in the masking algorithm.  The PSF convolved model image provides a good description of the complex, observed light distribution, making it possible to produce a deviation image (observed minus model divided by per pixel noise) where cosmic rays and other artifacts are visible at high statistical significance.}
\label{fig:singleepochimages}
\end{figure}

\subsection{Production of Model Single Epoch Image}
\label{sec:model}

To enable masking, accurate PSF and background models have to be extracted for each single epoch image.  These two actions are shown in the upper left portion of the data flow chart in Figure~\ref{fig:flowchart}.  The PSF model must follow the positional variations of the PSF within the single epoch image, because without that information it is not possible to use the coadd catalog to produce an accurate model for the single epoch image.  In CosmoDM we use {\tt PSFEx} to create a PSF model that is characteristically described by a series of $35\times35$~pixel kernels that are used in polynomial expansions in CCD $x$ and $y$ coordinates to track the positional variation of the PSF.  We typically allow for third order PSF variations along both dimensions.

A background model for the single epoch image is needed to enable comparison between the single epoch image and the model image, which we generate as a background free image.  There are various techniques for producing the background model, but within CosmoDM we use {\tt SExtractor} with the option  {\tt  -CHECKIMAGE\_TYPE -BACKGROUND}.  The background modeling parameters must be tuned to ensure that model background is accurate, or else the performance of the masking algorithm will be degraded.

To produce a high fidelity model single epoch image, we take as inputs the coadd catalog of model components extracted from the homogenized, median combine coadd and the PSF model for the single epoch image.  We use these together with {\tt Skymaker}~\citep{Skymaker09} to produce the model image.

To speed up the production of the model image, we exclude very faint objects from consideration that are not relevant for the masking.  Therefore, to produce the catalog input to skymaker we first select all objects from the coadd catalog having {\tt magerr\_model}$<0.3$, where {\tt magerr\_model} is the {\tt SExtractor} uncertainty on the model magnitude.  
For each remaining object we transform from sky location to single epoch image pixel coordinates using the world coordinate system (WCS) solution \citep{Greisen02} for the single epoch image.  To ensure that the impact of larger objects are properly tracked we include sources whose central pixel positions lie even somewhat outside the physical boundaries of the single epoch image.   In addition, we scale the magnitudes to account for the zeropoint difference between the coadd and single epoch image.  We then create a grid of PSF models for the single epoch image at 10 uniformly-spaced locations by running {\tt PSFex} together with the best fit PSF model for the single epoch image.  Using this catalog and the PSF model grid, we then simulate a background-free model image. 

We list below the parameters we use as input for simulations of stars or galaxies with {\tt Skymaker}.  We note that the coadd cataloging used for testing has been done by running {\tt SExtractor} with model fitting photometry with a {\tt Bulge+disk} model, but a sersic model is also available and could be used.  The list of these parameters are marked by the number of the column in which they appear in the input catalog.   The {\tt SExtractor} catalog parameter used for each column in the input catalog are also indicated.

\begin{itemize}
\item Col 1: Star/Galaxy Classification.  The star-galaxy classification is made  from the coadd catalog using the newly introduced star/galaxy classification parameter {\tt spread\_model}~\citep{Desai12,Bouy13}.  For the tests run here we adopt $\lvert {\tt spread\_model} \rvert < 0.002 $ in the coadd catalog to identify unresolved objects and model everything else as a galaxy.  When  the value of this  column is 200, {\tt Skymaker} does a bulge+disk modeling of the object.  Otherwise, if the value is 100  {\tt Skymaker}  models the object as a star.
\item Col 2, 3: Single epoch image $X$ and $Y$ position calculated from the sky positions of the object in the coadd image and the single epoch image WCS solution.
\item Col 4: Total Magnitude.  We use {\tt mag\_model} values from the coadd catalog after adjusting for the zeropoint difference between the coadd image and the single epoch image. 
\end{itemize}
The remaining quantities are included only for extended sources.
\begin{itemize}
\item Col 5: Bulge to total flux ratio {\tt FLUXRATIO\_SPHEROID}. 
\item Col 6: Bulge length scale (arcsec) {\tt 3600*SPHEROID\_REFF\_WORLD}.
\item Col 7: Bulge projected axis ratio {\tt SPHEROID\_ASPECT\_IMAGE}.
\item Col 8: Bulge position angle {\tt SPHEROID\_THETA\_IMAGE} (degrees). {\tt SPHEROID\_THETA\_IMAGE} must be transformed from the coadd image reference frame to the single epoch image reference frame.
\item Col 9: Disk length scale (arcsec) {\tt DISK\_THETA\_IMAGE}.
\item Col 10: Disk projected axis ratio {\tt DISK\_ASPECT\_IMAGE}.
\item Col 11: Disk position angle (degrees) {\tt DISK\_SCALE\_WORLD}.
\item Col 12: Redshift: set to zero. 
\end{itemize}

With the input catalog prepared as above and the PSF model, we then use {\tt Skymaker} 
%
%
to produce the output image, which is a 2048$\times$4096 model single epoch image.  This call includes the single epoch model image zeropoint and exposure time.  For our model image we have disabled the aureole and sky background and produce a noise free model.  Note that the {\tt Skymaker} parameter {\tt psf\_oversamp} is the inverse of the PSF model FITS keyword {\tt PSF\_SAMP}.  Figure~\ref{fig:singleepochimages} contains a portion of a model image created by {\tt Skymaker} in the middle panel, while the left panel contains the corresponding original background subtracted single epoch image.

\subsection{Masking Criteria}
\label{sec:masking}

Once we have created the model image,  we create a deviation image by subtracting the model image from the background subtracted single epoch image and scaling by the square root of the weight map image.  As noted above, the weight map image contains the inverse variance of the image noise, which contains the combination of the photon noise, read-out noise and the uncertainties in the various applied corrections.  Note that the weight map must include the contributions to the noise from the object flux or else there will be too many false detections within objects.  An example of a signal to noise deviation image is shown in the right panel of Figure~\ref{fig:singleepochimages}.

Any pixel in this deviation image is classified as a candidate artifact if it has a value $>$5, corresponding to a 5$\sigma$ positive deviation between the single epoch image and the underlying transient free model image.  This condition alone would lead to many false positives, because there is noise also in the model image.  More precisely, there are locations where the model fitting on the coadd image has failed and so the model light distribution is not a good representation of the transient free observed light distribution. 

To avoid these false positives we use the {\it coadd model residuals} image from the coadd cataloging (the third required input described in Section~\ref{sec:inputs}) to identify areas of the sky where the {\tt SExtractor} model fitting has failed.  Specifically, we use the WCS in the single epoch image along with {\tt SWarp} to produce a version of the model residuals image remapped to the reference frame of the single epoch image.  A candidate artifact pixel is vetoed if the model residuals image has a deviation with an absolute value greater than some threshold.  In our tests, we have found that an absolute value of 10 works well, corresponding to a deviation of the coadd model image from the median combine, PSF homogenized coadd that has 10$\sigma$ significance in either direction.

In many cases, bright stars are saturated and have long streaks along the readout direction that are called bleed trails.  These features are approximately persistent, and so they appear within the artifact free median combine coadd images.  Such features are not modeled by {\tt SExtractor} and therefore would trigger the model residuals image veto.  

Interestingly, we find that false positives are more likely around the saturated pixels of bright stars in locations where the coadd model residuals veto does not come into play.  Further testing is required to understand what appears to be additional noise in these regions of the image.  In the results presented below we have adopted a veto of all artifact candidates that lie within five pixels ($\sim1.25"$) of the saturated pixels in a bright star.

Finally, we find that often artifacts found in the first pass are surrounded by other pixels that are clearly contaminated but not at the level required to trigger the 5$\sigma$ threshold in the signal to noise deviation image.  This is in principle true for any imaged artifact, because the imaging PSF extends to large radius, spreading light from the artifact into the surrounding pixels. Thus, we pass through the image two more times examining the neighboring pixels of all previously marked artifact pixels.  In the second pass we examine all unmasked pixels within a two pixel distance of previously masked artifact pixels, marking those with signal to noise deviations greater than 2.5$\sigma$ as artifacts.  In a third pass we examine all neighboring pixels of artifacts, marking those with deviations greater than 1.5$\sigma$.  These second and third pass candidates are subjected to the same veto conditions as the first pass candidates.  This approach appears to work well in detecting the low surface brightness edges of artifacts like satellite trails and scattered light that have been imaged through the telescope optics.

The masking stage is finished as all transient artifacts are marked with a customized bit in the BPM.  Thereafter, the masked pixels are subjected to the cleaning process described in the next section.

\begin{figure}
\begin{center}
\includegraphics[width=0.8\columnwidth]{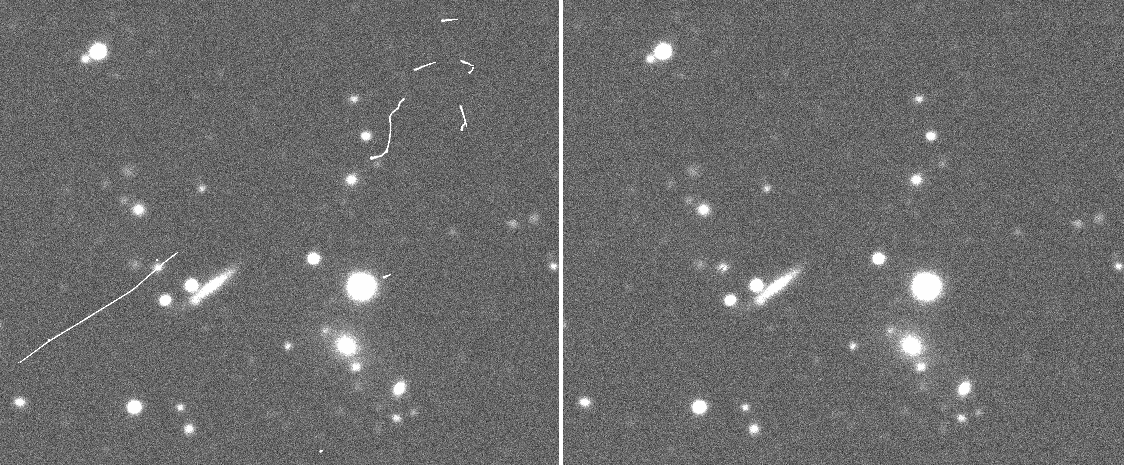}
\end{center}
\vskip-0.2in
\caption{DES-SV image (\#179097, CCD 27) cutout ($556 \times 463$ pixels) showing the original image with cosmic rays including one which intersects a star (left) and the image after masking and artifact removal (right).  Cosmic rays are identified and removed, even when they intersect real objects.}
\label{fig:craycutout}
\end{figure}

\section{Algorithm for Removal of Artifacts}
\label{sec:removal}

There are situations in the construction of coadd images where it is beneficial to remove identified artifacts in the underlying single epoch images to avoid their contaminating neighboring locations on the sky.  Moreover, in certain circumstances artifacts can be removed with high accuracy, resulting in only a small increase of noise over the case where the artifact had not been present.    

In the process of building a coadd image from a stack of single epoch images, the single epoch images are all remapped into the pixel reference frame of the coadd image.  Because of the physical imperfections of the detector system and the optical distortions of the camera and telescope optics, the single epoch image pixel reference frames are characteristically distorted tangent plane projections.  While in principle one has the freedom to adopt any reference frame for the coadd image, it is often the case that these are described by perfect tangent plane projections.  This remapping of the single epoch images is an interpolation that involves the values of the single epoch image pixels that lie closest to the sky position of the coadd pixel.  Generically, then, depending on the size of the interpolation kernel, a single pixel artifact in a single epoch image will lead to the contamination of an island of surrounding pixels in the coadd.  In the case that the coaddition algorithm is set to ignore all pixels that are contaminated by artifacts this could lead to information loss, 
whereas in cases where only those pixels contaminated above some threshold are ignored the resulting light distribution in the coadd could be affected by some residual contamination from the single epoch image artifact.

In the case of PSF homogenization the single epoch images are all brought to a common PSF and a common pixel reference frame as part of the coadd construction.  In this case the spatially varying smoothing kernel that is used for homogenization has characteristically the extent of the PSF.  If one adopts fast Fourier transform methods to enable an efficient homogenization, a single pixel artifact then impacts an island of surrounding pixels of scale similar to the PSF.  Within CosmoDM we employ fast Fourier transform convolution methods and remap with a Lanczos kernel; therefore, we seek to remove artifacts where possible. 

\begin{figure}
\begin{center}
\includegraphics[width=0.6\columnwidth]{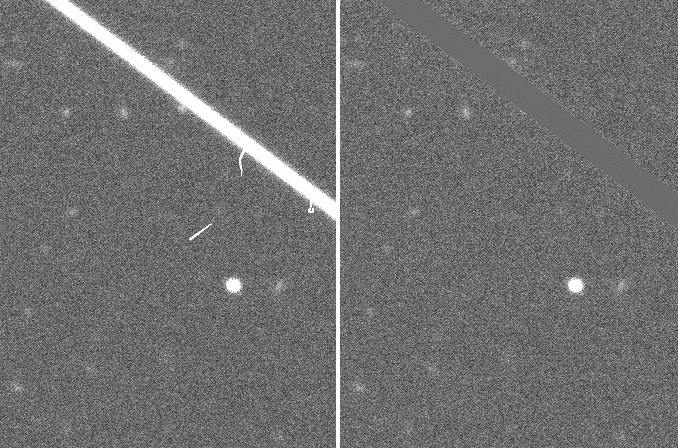}
\end{center}
\vskip-0.2in
\caption{DES-SV image (\#166255, CCD 60) cutout ($670 \times 450$ pixels) showing the  original image with the satellite trail and cosmic rays  (left) and the masked and artifact free image (right).  Within the satellite trail, which is an imaged artifact, we do not do a PSF based interpolation and instead replace the pixel values with the local background.  The cosmic rays, on the other hand, are non-imaged artifacts and are removed using PSF interpolation.}
\label{fig:satellitetrail}
\end{figure}

\subsection{Modeling the Light Distribution}

The observed light distribution in an image is the convolution of the underlying, true light distribution with the PSF after the addition of Poisson noise and noise in the detrending corrections.  For DECam imaging where the characteristic FWHM of the PSF is just under 1~arcsec and the pixel size of the detectors is 0.263~arcsec,  there is significant redundant information among neighboring pixels for each imaged source.
Within this context, a non-imaged artifact-- that is, an artifact like a cosmic ray or a bad column that has not been imaged through the telescope and camera optics--  can be effectively replaced through interpolation with only a modest increase in noise over the case where the artifact was not present.  This applies to all non-imaged artifacts that have widths that are small compared to the FWHM of the PSF.  

Imaged artifacts like satellite trails have sizes that are as large as or larger than the PSF, and therefore it is not possible to interpolate over them without  adopting a strong prior on the underlying light distribution.  For example, one could adopt a model that there is only sky beneath the satellite trail, but then the accuracy of the removal hinges on this assumption.  In many circumstances the probability that this model is correct is quite good, but in crowded areas on the sky the probability falls, making such an approach unwarranted.  Within the context of building coadds where there are multiple single epoch exposures in each part of the sky, the safest approach is to mask all these affected pixels, set their weights to zero and not allow them to contribute to the coadd.

Our approach to interpolation uses the local PSF as constrained using {\tt PSFEx}.  Because the observed or imaged light distribution $I$ is the convolution of the underlying light distribution $M$ with the PSF $P$ with the addition of appropriate Poisson noise and other noise coming from detrending corrections $N$, 
\begin{equation}
I=M\otimes P + N
\end{equation}
a straightforward method for determining the proper values of contaminated pixels would be through a forward modeling process.  One would iteratively build up the best fit underlying model using the uncontaminated surrounding pixels to fix, for example, the morphological and flux characteristics of the observed light distribution.  Again, in a situation where the artifact is non-imaged and small in width compared to the PSF this is a high fidelity process and will work for the replacement of contaminated pixels whether they lie in background regions or in either unresolved or resolved objects.  

This forward modeling process is exactly what {\tt SExtractor} currently does.  For simplicity and to remove sensitivity to the fidelity of the underlying model light distribution $M$ we use direct interpolation with the local PSF model as an interpolation kernel on the observed light distribution in most circumstances.  That is, the replacement value $\tilde I(x,y)$ of the masked pixel at location $(x,y)$ is just the convolution of the masked image $I$ with the PSF $P$ evaluated at that single location 
\begin{equation}
\tilde I(x,y) = I\otimes P\, |_{\,(x,y)}
\end{equation}
This approach requires a simple PSF weighted sum over the neighboring, unmasked pixels in the image and is therefore straightforward and efficient.  In the limit where the background sources are faint (i.e. absence of source or a typical faint star or galaxy) the results of the direct interpolation and the forward modeling do not differ significantly, but for bright sources where the fractional noise per pixel is very low this approach is inadequate.  Thus, for these bright sources we use forward modeling for the replacement of contaminated pixels.

\begin{figure}
\begin{center}
\includegraphics[width=0.6\columnwidth]{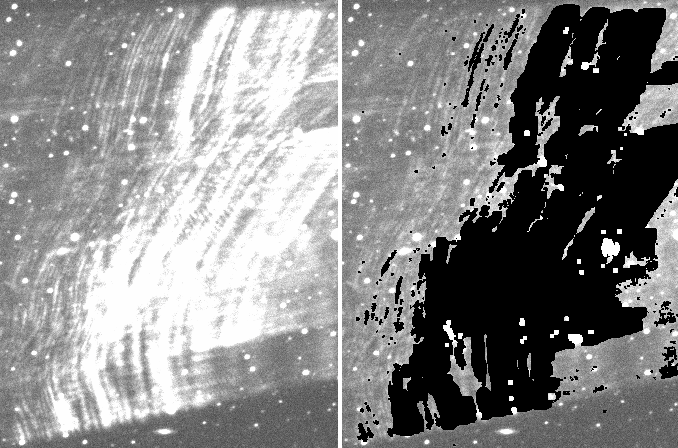}
\end{center}
\vskip-0.2in
\caption{DES-SV image (\#158781, CCD 62) cutout ($670 \times 900$ pixels) showing the original image with the scattered light (left), and the masked image with all masked pixels set to black (right).  A large area of scattered light has been masked by our algorithm. The unmasked pixels in in the scattered light area are pixels that have been flagged as artifacts because they are saturated cores of bright stars.}
\label{fig:scatteredlight}
\end{figure}

\subsection{Interpolation over Non-imaged Artifacts}
\label{sec:interpolation}

For direct interpolation to determine the value of a masked pixel we center the local model of the PSF at that pixel and first determine what fraction $f_{umsk}$ of the PSF probability distribution overlaps unmasked pixels.  If $f_{umsk}$ lies above some threshold fraction $f_{th}$, we then proceed with the interpolation.  Through interpolation we estimate the value $\tilde I(x,y)$ of the masked central pixel as a PSF weighted sum over the unmasked pixels within the interpolation kernel, accounting properly for the portion of the kernel that overlaps masked pixels in the image.  In addition, we calculate the associated inverse variance weight $\sigma^{-2}(x,y)$ of the  interpolated pixel using Gaussian error propagation over the unweighted pixels within the PSF kernel.  Finally, we sample the expected observed value $\tilde I$ of the pixel as a Gaussian random deviate with the measured mean $I$ and variance $\sigma^2$ so that the pixel to pixel variations of the interpolated pixels are comparable to those of the surrounding uncontaminated pixels.  For each interpolated pixel a bit is set in the BPM image.

In the tests we describe below on DECam data the PSF kernel is $53\times53$ pixels, and the minimum unmasked kernel threshold probability is $f_{th}=0.1$.  With these values we find good performance for the pixel interpolation.  Those pixels that do not meet the threshold probability are replaced with an estimate of the sky brightness at that location (taken from our background model for the single epoch image) and their weight map pixels are set to zero.  Note that this approach allows for a few pixels of interpolation around the edges of larger, imaged artifacts like satellite trails and scattered light, but the bulk of the pixels affected by imaged transients are simply replaced by sky background values with their weight maps set to zero, ensuring that none of these masked pixels are used in the coadd.

\subsection{Bright Star Replacement}
\label{sec:replacement}

We replace the saturated pixels from bright stars and their associated bleed trails with values of the underlying, best fit model light distribution produced by {\tt SExtractor}.  For all saturated objects we examine a rectangular region centered on the object that has a narrow extent equal to $5 \times {\tt FLUX\_RADIUS}$ and an extent along the perpendicular axis that is increased until no more saturated pixels are found.  Within this rectangle all saturated pixels are placed with the value of the underlying PSF corrected model plus the value of the background model at that location.  This essentially means that bleed trails extending far from the stellar core are replaced by the sky, whereas saturated central pixels are replaced by the best fit model of the object.  As with interpolation, a bit in the BPM image is set to record that these pixel values arise from the model image.  The new weight map value of each replaced pixel in the saturated stellar core is scaled to 1\% of the original weight map value.  The bleed trail pixels whose values are replaced with sky values have weight map pixels set to zero.

\section{Application to Dark Energy Survey Science Verification Data}
\label{sec:application}

In this section we describe a test dataset and present examples and statistics for the algorithm as applied to these data.  We end with a discussion of how the masking impacts the depth of the coadd images by allowing one to use inverse variance weighted coadds rather than median combine coadds.

\subsection{Test Dataset}

To test and tune the tools described above we process the Dark Energy
Survey~\citep[DES;][]{DES05,DES16} Science Verification (SV) data.  The DES is conducted using  a 570~Mpixel Dark Energy Camera~\citep{Decam15} with a $3^{\circ}$~deg$^2$ 
field of view, which was commissioned in Sept. 2012. Although the DES officially began in August 2013, there was a science verification (SV) phase prior to that, during which  extensive verification of the camera and associated operations took place.
During SV, about 130 deg$^2$ of the southern sky overlapping the South Pole Telescope survey~\citep{SPT11} with  $ 60^{\circ}<\alpha< 90^{\circ} $ and $-65^{\circ} < \delta < -45^{\circ}$ was imaged at full DES depth in the $grizY$ bands, corresponding to 10 imaging layers in each band at a given point.  

For the testing we created 10 coadd tiles, each with solid angle $\sim$1~deg$^2$, in $griz$ bands using CosmoDM.  Our input data were the publicly available raw data.  The data quality for these tiles is very good. The median stellar locus scatter when doing absolute photometric calibration is 31 and 16 mmag in the $g-r$ versus $r-i$ and the $r-i$ versus $i-z$ color-color spaces, respectively. We also find that the median single epoch photometry in the repeatability plots  at the bright end around 14-15 mag is between 7-8 mmag in all four bands \citep[see][for numbers from other surveys]{Desai12,Liu15}.

The locations of these test tiles were chosen to span both typical extragalactic fields as well as high stellar density fields nearer to the LMC.  In the tile lists that appear in the tables below, these high stellar density fields have lower declinations ($\delta<-61^\circ$).

\subsection{Masking and Removal Examples}
We present some image cutouts taken from our test dataset to demonstrate that our algorithm successfully identifies the different types of transients that are seen in astronomical images.  Figure~\ref{fig:craycutout} contains an example of an image with cosmic rays and the automatically masked and cleaned image.  As is clear, all the cosmic rays including a particularly long one that intersects a star are identified and replaced with estimates of the underlying, uncontaminated flux. As described above, in cases of non-imaged artifacts like cosmic rays it is generally possible to replace contaminated pixels without bias and with only a small enhancement of the noise over what the pixel would have had if the cosmic ray had not been present. 

In contrast, Figure~\ref{fig:satellitetrail} contains an example of an image with a satellite trail.  Our method has no trouble identifying these artifacts, but for imaged artifacts it is not possible to replace the contaminated pixels, given that the artifact is as large as or larger than a PSF.  Thus, in this figure we have simply replaced the contaminated pixels with the sky and have set the weight map values for those pixels to zero to ensure that they do not contaminate the coadd to which they contribute. 

In Figure~\ref{fig:scatteredlight}, we show an example of an image that is affected by scattered light.  This kind of large scale feature can also be identified automatically by our method.  Given the scale of the image artifact, there is not enough information to allow removal or replacement with local interpolation, so we simply mask these pixels and set them to the background value.  In the right panel of Figure~\ref{fig:scatteredlight}, all pixels flagged as scattered light are shown in black for illustration.  Our algorithm flags all  the scattered light induced pixels, which are not already flagged. The only non-detected scattered light pixels for this image are due to saturated stars which also fall in the midst of the scattered light.  By using a grow radius with multiple applications it is possible to aggressively remove such large features, ensuring no measurable contamination remains or to be less aggressive, resulting in slightly contaminated, unmasked pixels around the edge of the feature.

\begin{figure}
\vskip-0.1in
\begin{center}
\includegraphics[width=0.33\columnwidth]{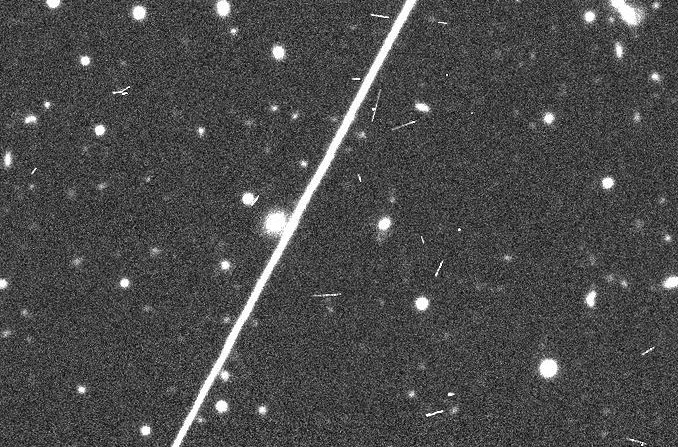}
\includegraphics[width=0.33\columnwidth]{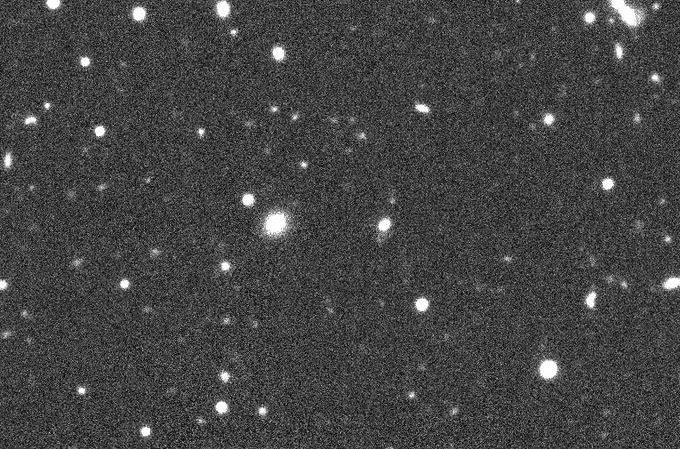}
\includegraphics[width=0.33\columnwidth]{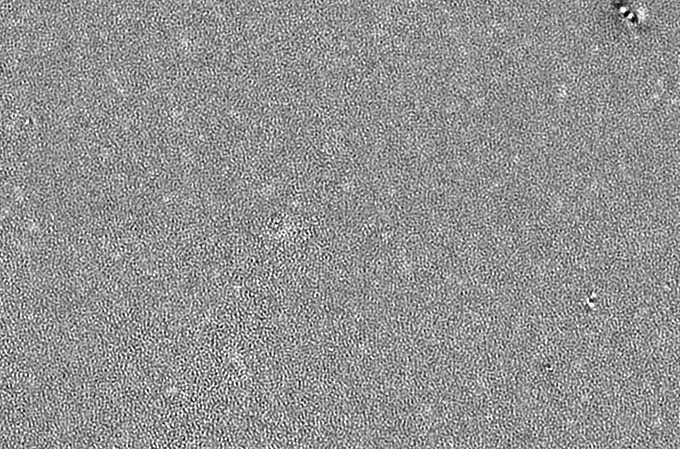}
\end{center}
\vskip-0.2in
\caption{DES-SV coadd image 0427-4539 cutout ($680 \times 450$ pixels) produced with inverse variance weighted combine of single epoch images prior to masking (left) and after masking (middle).  All the cosmic rays and the satellite trail have been effectively masked and removed.  The right panel contains the {\tt SExtractor} $-models$ image for the PSF homogenized coadd.  This image shows mostly noise with the exception of the bright core regions of asymmetric objects (upper right portion of image), which are not well fit.}
\label{fig:coadd}
\end{figure}

In Figure~\ref{fig:coadd}, we also show the impact of our masking algorithm on one of our coadd test tiles.  We show the coadd created using inverse variance weighted mean combine with unmasked single epoch image inputs (left), masked single epoch image inputs (middle) and an {\tt SExtractor} $-models$ image (right) that highlights how well the model fitting works in describing the light distribution once the artifacts have been removed.   As can be seen, the cosmic ray contamination, which is quite severe in a stack of ten exposures, has disappeared.

We have visually scanned  the ten coadded images created using the masked single epoch images and compared them to the median combine coadds to look for missed artifacts. The only unmasked cosmic rays we find are in the vicinity of very bright stars for which model-fitting fails and the pixels in the $-models$ image exceed the $10\sigma$ threshold.  The fraction of pixels on the sky that cannot be masked due to model fitting failures with the current version of {\tt SExtractor} is less than 0.5~percent.  We present those numbers in the next section.

\begin{table*}
\begin{center}
\begin{small}
\begin{tabular}{|l | rr | lrr | rr | lrr | lrr | lrr}
\hline
 \multicolumn{1}{l}{} & \multicolumn{2}{c}{Non-Imaged}   &
\multicolumn{3}{c}{Imaged} & \multicolumn{2}{c}{Saturated} & \multicolumn{3}{c}{False Positives} & \multicolumn{3}{c}{On Objects} \\
 \multicolumn{1}{l}{Tilename} &  \multicolumn{1}{c}{$N_p$} &\multicolumn{1}{c}{$N_A$}  &\multicolumn{1}{c}{$f_{im}$} & \multicolumn{1}{c}{$N_p$} &\multicolumn{1}{c}{$N_A$}  & \multicolumn{1}{c}{$N_p$} & \multicolumn{1}{c}{$N_A$} &  \multicolumn{1}{c}{$f_{im}$} & \multicolumn{1}{c}{$N_p$} & \multicolumn{1}{l}{$N_A$} &  \multicolumn{1}{c}{$f_{im}$} & \multicolumn{1}{c}{$N_p$} & \multicolumn{1}{c}{$N_A$}   \\
\hline

0423-5236 & 2143 & 63 & 0.9 & 10336 & 1 & 6912 & 13  & 77 & 51 & 3 & 83 & 32 & 4\\
0427-4539 & 2268 & 67 & 0.9 & 23233 & 1 & 6594 & 11 & 86 & 77 & 5 & 86 & 84 & 5\\
0448-5634 & 2065 & 65  & 0.5 & 18964 & 1 & 9733 & 18 & 82 & 79 & 5 & 85 & 32 & 4\\
0454-6032  & 2133 & 65 & 0.6 & 24449 & 1 & 10037 & 21 & 79 & 46 & 5 & 86 & 213 & 5\\
0455-4937 & 2386 & 70 & 0.6 & 21508 & 1  & 7962 & 15  & 91 & 101 & 6 & 88 & 35 & 4\\
0455-5833 & 2283 & 65 & 0.8 & 8298 & 1 & 10698 & 19 & 81 & 58 & 4  & 86 & 206 & 5\\
0502-5335  & 2287 & 64 & 0.3 & 33765 & 1 & 7578 & 14 & 86 & 102 & 6  & 83 & 29 & 4\\
0503-6231 & 2737 & 89 & 1.0 & 10990 & 1 & 11459 & 19  & 84 & 140 & 7 & 82 & 34 & 4\\
0511-6231 & 2960 & 104 & 0.4 & 13060 & 1 & 10792 & 18 & 85 & 122 & 8 & 95 & 136 & 19\\ 
0538-6330 & 4430 & 166 & 0.3 & 7378 & 1  & 10261 & 20  & 88 & 209 & 11 & 96 & 336 & 48\\
\hline
\end{tabular}
\caption{Statistics on the number of transient artifacts-- grouped into non-imaged (e.g. cosmic rays), imaged (e.g. satellite trails, scattered light), saturated pixels, false positives, and non-imaged artifacts overlapping real objects-- per single epoch image (each CCD) within each of the ten test tiles.  The first column is the tilename, and then for each type of artifact we present the percentage of images  $f_{im}$ containing at least one artifact, the median number of masked pixels per image $N_p$, and the median number of distinct artifacts per image $N_A$ for images that contain at least one artifact.  Note that $f_{im}$ for non-imaged and saturated artifacts is one for all tiles and hence is not shown.}\label{tab:statistics}
\end{small}
\end{center}
\end{table*}

\subsection{Masking and Removal Statistics}
\label{sec:stat}
After we run our composite masking code on 10 coadd tiles (with each tile containing four bands), we then count the number of imaged and non-imaged artifacts as well as the total number of pixels affected in each image.  We find the total number of distinct artifacts in an image using an algorithm developed first for studying the X-ray isophotal sizes of galaxy clusters \citep{Mohr97}.   The algorithm shares some characteristics with the {\tt K-Means} clustering algorithm~\citep{Babu}.   All pixels, which are contiguously connected are hierarchically grouped into distinct clusters, enabling the measurement of characteristics such as the total number of pixels. Table~\ref{tab:statistics} contains these statistics on a per image basis.   The statistics are grouped from left to right into different types of artifacts:  (1) non-imaged artifacts, (2) imaged artifacts, (3) saturation artifacts, (4) false positives and (5) artifacts that overlap real sky objects.  By false positives, we mean the number of  incorrectly identified transient defects. These false positives are identified by looking for overlap of our transient defects with 2MASS objects in each image.

For each coadd tile, we indicate the percentage $f_{im}$ of single epoch images that have at least one artifact of each type, the median number of masked pixels $N_p$ per image, and the median number of  distinct, masked artifacts per image that contains at least one artifact.  In the case of non-imaged artifacts and saturated objects, essentially every image contains at least one of these so we do not present $f_{im}$.

We detect between 60 and 70 non-imaged artifacts (e.g., cosmic rays) per image, and these artifacts typically contaminate about 2200 pixels per image. Note that the tiles near the bottom of the table that have high stellar densities show a systematically higher number of non-imaged artifacts, and this is likely a reflection of increased false positives when the stellar density is high.  Less than 1~percent of the images have an imaged artifact (e.g., satellite trails, scattered light), and these are rare enough that we typically see no more than one on an affected image.  Typically, these imaged artifacts affect 10,000 to 20,000 pixels. The median number of saturated bright stars is about 15 per image, affecting 7000 to 10,000 pixels.  About 85~percent of images contain false positives, which include both variable stars and false objects associated with structures around bright, saturated stars.  The typical number of pixels per false positive is between 15 and 20, and there are typically five of these per image.  As expected, these false positive rates are higher where the stellar density is higher in the most southern tiles that are nearer to the LMC.

We also count the number of cosmic rays intersecting an object, and these numbers appear in the last column of Table~\ref{tab:statistics}.  To calculate these statistics, we created a segmentation image by running {\tt SExtractor} on the masked image. We then count the non-imaged artifacts which intersect a real object.  About 85~percent of images contain at least one non-imaged artifact that intersects a real object, and most images contain about four or five . 

\begin{table*}
\begin{center}
\begin{small}
\begin{tabular}{|l|rr|rr|rr|rr|}
\hline
\multicolumn{1}{c}{} & \multicolumn{2}{c}{$g$} & \multicolumn{2}{c}{$r$} & \multicolumn{2}{c}{$i$} & \multicolumn{2}{c}{$z$}  \\
 \multicolumn{1}{l}{Tilename} & \multicolumn{1}{c}{$m_{5\sigma}$} & \multicolumn{1}{c}{$\Delta m_{5\sigma}$} &  \multicolumn{1}{c}{$m_{5\sigma}$} & \multicolumn{1}{c}{$\Delta m_{5\sigma}$} &  \multicolumn{1}{c}{$m_{5\sigma}$} & \multicolumn{1}{c}{$\Delta m_{5\sigma}$} & \multicolumn{1}{c}{$m_{5\sigma}$} & \multicolumn{1}{c}{$\Delta m_{5\sigma}$} \\
\hline
0423-5236 &  24.74 & 0.09 & 24.49  & 0.10 & 23.69 & 0.14  & 23.0  & 0.10  \\
0427-4539 & 24.59 & 0.03 & 24.21 & 0.05 & 23.50 & 0.08 & 22.87 & 0.08  \\
0448-5634 & 24.69 & 0.07 & 24.46 & 0.06 & 23.90 & 0.05 & 23.22 & 0.14 \\
0454-6032 & 24.77 & 0.12 & 24.43 & 0.09 & 23.67 & 0.06 & 22.97 &  0.11 \\
0455-4937 & 24.01 & 0.26 & 23.81 & 0.27 & 22.99 & 0.27 & 22.25 & 0.25 \\
0455-5833 & 24.74 & 0.09 & 24.30 & 0.11 & 23.81 & 0.11 & 23.2 & 0.11 \\
0502-5335 & 24.55 & 0.24 & 24.21 & 0.25 & 23.51 & 0.08 & 22.91 & 0.09 \\
0503-6231 & 24.61 & 0.10  & 24.22 & 0.13  & 23.81 & 0.19 & 23.09 & 0.07  \\
0511-6231 & 24.48 & 0.09 & 24.13 & 0.16 & 23.59 & 0.10 & 22.92 & 0.14 \\
0538-6330 & 24.47 & 0.09 & 23.97 & 0.11 & 23.16 & 0.16  & 22.45  & 0.10  \\
\hline
\end{tabular}
\caption{$5\sigma$ depths $m_{5\sigma}$ in 3'' diameter apertures using measured sky noise for inverse variance weighted mean combine coadds along with the difference in depth $\Delta m_{5\sigma}$ between these weighted combine and the median combine coadds.}\label{tab:depthcomparison}
\end{small}
\end{center}
\end{table*}

\subsection{Masking Impact on Coadd Depth} 
After masking is complete on our single epoch test dataset, we rebuild the coadds using the inverse variance weighted mean combine option in {\tt SWarp}.   Here we compare the depths of these images to the depths of the median combine coadd images.  For this purpose we extract 5$\sigma$ depths in 3'' diameter apertures.  To calculate the depths, we measure the sky noise variance $\sigma_{sky}$ in 10,000 random apertures throughout the image which are not contaminated by any object pixels as identified using the {\tt SExtractor} segmentation image.  The 5$\sigma$ depth $m_{5\sigma}$ is then $ZP -2.5log_{10}(\sigma_{sky})$.  Table~\ref{tab:depthcomparison} contains the measured depths $m_{5\sigma}$ for the weighted mean combine coadds and the offset relative to the median combine coadds. We find that the weighted combine coadds are deeper by $\sim$0.1 to 0.2~mag.  In the limit of identical noise and no losses to masking we would expect the inverse variance weighted mean combine images to be $\sim$0.25~mag deeper \citep{Kendall77}.

\section {Discussion of Failures and Future Improvements}
\label{sec:discussion}

As previously noted, we find that  our masking algorithm only misses the artifacts due to {\tt SExtractor} model-fitting failures in the vicinity of very bright stars and also in the cores of nearby galaxies, because pixels that are poorly fit (deviation between model and observation that exceed 10$\sigma$ significance) cannot be masked with our approach.  Thus,  our masking algorithm is dependent on the fidelity of model-fitting photometry in {\tt SExtractor}, which continues to be improved by Dr. Bertin and collaborators.  

For each of the 10 coadd tiles we investigate the statistics of model-fitting failures by measuring the number of  pixels from all objects in the coadd catalog extracted with {\tt Bulge+Disk} model-fitting that exceed the $10\sigma$ threshold in the $-models$ image. For every coadd image, we created a segmentation image from {\tt SExtractor}, which maps the list of pixels associated with a given object. For these pixels, we then count the number of pixels in the $-models$ image, whose absolute value exceeds  the $10\sigma$ statistical threshold.  Because model-fitting failures occur mainly at bright magnitudes, we restrict the statistics to all objects with magnitude error $<0.1$~mag.

Tab.~\ref{tab:mfpfailurestats} contains these statistics for model-fitting failures broken down by imaging band $griz$, presenting for each the fraction of detected objects $f_{obj}$ with model fitting failures, the fraction of object pixels $f_{opix}$ within those failed objects that have deviations greater than 10$\sigma$ significance, and the fraction of image pixels $f_{ipix}$ that have model fitting failures.  For most tiles, between 2 and 10\% of objects contain at least one pixel which crosses the 10$\sigma$ threshold. The typical fraction of failed pixels per object that contains at least one failed pixel is about 5\%.  And the total number of failed pixels in the coadd image is typically less than 0.5\%.  For all these cases the fraction is higher for the tile closest to the LMC (0538-6330), because of its very high stellar density. 

\begin{table*}
\begin{center}
\begin{small}
\begin{tabular}{|l|rrr|rrr|rrr|rrr|}
\hline
\multicolumn{1}{c}{} & \multicolumn{3}{c}{coadd $g$ band} & \multicolumn{3}{c}{coadd $r$ band} & \multicolumn{3}{c}{coadd $i$ band} & \multicolumn{3}{c}{coadd $z$ band}  \\
\multicolumn{1}{l}{Tilename}  & \multicolumn{1}{c}{$f_{obj}$} & \multicolumn{1}{c}{$f_{opix}$} &  \multicolumn{1}{c}{$f_{ipix}$} &  \multicolumn{1}{c}{$f_{obj}$} & \multicolumn{1}{c}{$f_{opix}$} &  \multicolumn{1}{c}{$f_{ipix}$} &  \multicolumn{1}{c}{$f_{obj}$} & \multicolumn{1}{c}{$f_{opix}$} &  \multicolumn{1}{c}{$f_{ipix}$} & \multicolumn{1}{c}{$f_{obj}$} & \multicolumn{1}{c}{$f_{opix}$} &  \multicolumn{1}{c}{$f_{ipix}$} \\
\hline
0423-5236 & 6.30 & 3.16 & 0.16 & 7.63 & 3.16  & 0.19  & 4.73 & 2.53 & 0.16 & 3.33 & 1.44 & 0.08 \\
0427-4539 & 1.19 & 3.13 & 0.03 & 1.52 & 2.67 & 0.03 & 0.73 & 1.56 & 0.01  & 0.46 & 0.75 & 0.01 \\
0448-5634 & 6.24 & 6.59 & 0.33 & 6.53 & 6.56 & 0.36 & 4.98 & 5.22 & 0.29 & 4.33 & 4.51 & 0.23 \\ 
0454-6032 & 7.49 & 8.34 & 0.41 & 6.85 & 8.48 & 0.41 & 4.97 & 5.41 & 0.27 & 4.26 & 4.92 & 0.23 \\  
0455-4937 & 6.01 & 4.01 & 0.16 & 5.88 & 3.67 & 0.17 & 4.09 & 1.98 & 0.09 & 5.71 & 2.83 & 0.12 \\
0455-5833 & 8.64 & 8.89 & 0.44 & 7.88 & 7.90 & 0.43 & 5.51 & 4.97 & 0.27 & 5.41 & 5.84 & 0.30 \\
0502-5335 & 9.06  & 9.94 & 0.36 & 8.90 & 7.32 & 0.29 & 5.41 & 2.83 & 0.11 & 5.57 & 4.65 & 0.18 \\
0503-6231 & 7.40 & 3.62 & 0.36  & 5.92 & 3.32 & 0.34 & 3.90  & 2.48 & 0.26 & 2.74 & 2.62 & 0.26 \\
0511-6231 & 10.14  & 4.48 & 0.52  & 7.63  & 3.10 & 0.38 & 4.36 & 2.15 & 0.26 & 2.73 & 1.69 & 0.20 \\
0538-6330 & 29.23 & 7.97 & 1.77 & 25.4 & 5.71 & 1.29 & 15.77 & 4.52 & 0.99 & 5.42 & 1.05 & 0.24 \\
\hline
\end{tabular}
\caption{Statistics of model-fitting failures for the ten processed tiles in bands $griz$, where we denote model-fitting failures as pixels in the model residuals image showing $>$$10\sigma$ deviations between model and observation. We present the percentage of objects with at least one failed pixel $f_{obj}$, the average percentage of object pixels with model-fitting failures $f_{opix}$ within objects that have at least one pixel that has failed, and the percentage of pixels with model fitting failures in the entire image $f_{ipix}$.  Note that these statistics are obtained for objects with {\tt magerr\_model}$<$0.1.}
\label{tab:mfpfailurestats}
\end{small}
\end{center}
\end{table*}

In summary, typically more than 99.5\% of the image pixels within the DES-SV test dataset can be modeled adequately by {\tt SExtractor} to enable the masking algorithm presented here to work.  As further improvements are made to the model fitting this fraction will increase.

\section{Conclusions}
\label{sec:conclusions}

We have developed an automated masking algorithm for the detection of all transient artifacts typically found in astronomical images.  These artifacts include non-imaged artifacts such as cosmic rays as well as imaged artifacts like satellite trails and scattered light.  This algorithm uses the multiple images of each portion of the sky to build a model of the non-varying or persistent light distribution of both resolved and unresolved sources, and therefore a requirement for this algorithm is that the dataset include multiple images of each sky location.

The algorithm builds the persistent light distribution through the median combination of PSF homogenized single epoch images into a coadd image with a well behaved (typically spatially invariant) PSF.  Then PSF corrected model fitting to the light distribution is carried out using {\tt SExtractor}.  The resulting coadd catalog is used together with the single epoch image PSF model to construct a single epoch model image.  Deviations between the observed light distribution in the single epoch image and the model image then indicate pixels that are affected by artifacts.  These artifacts can be identified whether or not they overlap real objects.  A small fraction of the image pixels ($<0.5$\%) cannot be used to identify artifacts because of failures in the model fitting to the PSF homogenized coadd images.  Thus, our algorithm automatically identifies and masks artifacts in $>99.5$\% of the single epoch image pixels.

We discuss the removal of artifacts that are smaller than the PSF, noting that for these non-imaged artifacts like cosmic rays, it is possible to predict the uncontaminated pixel values with high accuracy and only a small increase in noise as compared to the case where there had been no artifact.  For imaged artifacts that are as large or larger than the PSF, accurate removal is not possible without prior knowledge of the uncontaminated light distribution.

We apply our algorithm to ten $\sim1$~deg$^2$ coadd tiles from the DES science verification dataset, where each tile has been imaged approximately 10 times in each band $griz$.  The algorithm performs well, masking and replacing 60 to 70 cosmic rays in every single epoch image (approximately $10'\times20'$ regions of the sky).  In addition, there are typically one or two imaged artifacts identified and masked on each image as well as approximately 15 saturated stars whose saturated pixels can be replaced by a model constrained using the unsaturated pixels.  False positives, which are associated with time variable objects, are found at the rate of about 5 per image, and-- as expected-- the false positive rate increases in tiles with higher stellar density.  

We recreate all ten coadd tiles using the masked single epoch images, demonstrating that all transient artifacts have disappeared and enabling a dramatic improvement in aesthetic image quality.  Creating the coadd images using inverse variance weighted mean combine results in an increase in depth of about 0.1 to 0.2~mag in comparison to the original median combine coadds.   

\section*{Acknowledgements}

We acknowledge the Deutsches Zentrum f\"ur Luft- und Raumfahrt (DLR) support of the LMU Euclid program (50QE1201) and the support by the DFG Cluster of Excellence ``Origin and Structure of the Universe'' and the associated Computational Center for Particle and Astrophysics (C2PAP).  The data processing has been carried out on the computing facilities of the C2PAP, which are operated by the Leibniz Rechenzentrum (LRZ).

We acknowledge the Euclid Consortium, the European Space Agency and the support of a number of agencies and institutes that have supported the development of Euclid. A detailed complete list is available on the Euclid web site (http://www.euclid-ec.org ). In particular the Agenzia Spaziale Italiana, the Centre National dÕEtudes Spatiales, the Deutches Zentrum fur Luft- and Raumfahrt, the Danish Space Research Institute, the Funda‹o para a Cinca e a Tecnologia, the Ministerio de Economia y Competitividad, the National Aeronautics and Space Administration, the Netherlandse Onderzoekschool Voor Astronomie, the Norvegian Space Center, the Romanian Space Agency, the United Kingdom Space Agency and the University of Helsinki.

This project used public archival data from DES. Funding for the DES Projects has been provided by the U.S. Department of Energy, the U.S. National Science Foundation, the Ministry of Science and Education of Spain, the Science and Technology Facilities Council of the United Kingdom, the Higher Education Funding Council for England, the National Center for Supercomputing Applications at the University of Illinois at Urbana-Champaign, the Kavli Institute of Cosmological Physics at the University of Chicago, the Center for Cosmology and Astro-Particle Physics at the Ohio State University, the Mitchell Institute for Fundamental Physics and Astronomy at Texas A\&M University, Financiadora de Estudos e Projetos, Fundacao Carlos Chagas Filho de Amparo a Pesquisa do Estado do Rio de Janeiro, Conselho Nacional de Desenvolvimento Cient\'ifico e Tecnol\'ogico and the Minist\'erio da Ci\^encia, Tecnologia e Inovacao, the Deutsche Forschungsgemeinschaft and the Collaborating Institutions in the Dark Energy Survey.  

\bibliographystyle{model2-names}
\bibliography{paper}

\end{document}